ON ORDER AND RANDOMNESS:  A VIEW FROM THE EDGE OF CHAOS
Gennady Shkliarevsky
Bard College

"[QM] yields much, but it hardly brings us closer to the Old One's secrets.  I, in any case, am convinced that He does not play dice."

--A. Einstein

"Not only does God definitely play dice, but He sometimes confuses us by throwing them where they can't be seen."

--S. Hawking

Introduction

Production of knowledge is undoubtedly one of the most defining features of human civilization.  We take enormous pride in how much we know and how much we can control the world in which we live.  We also greatly depend on the production of knowledge in ensuring our prosperity and wellbeing, safety and security, our very future.  Indeed, it is no exaggeration to say that our very sense of dignity and self-worth is due in no small degree to our unique capacity to generate knowledge.

There is a close relationship between production of knowledge and critical introspection into the ways we know.  Our mind does not have direct access to reality and all our knowledge is mediated.  The fact that we understand reality through the prism of our mind and its constructs implies that what we know critically depends on the tools we use, including our mental tools.  Every advance in our knowledge has always involved critical rethinking of the way we know.  What we know about how we know and how well we control our mental tools has a decisive effect on our knowledge production.  Without such critical introspection our knowledge can stagnate and become obsolete.

The main thrust of this paper is to show the relationship between how we know and what we know.  In arguing this point, the paper will look at one problem that profoundly divides our scientific community today.  It concerns the nature of reality.  There are currently two positions on this issue.  One position asserts that at its most fundamental level reality is acausal, random and unpredictable, and that statistical probability is the most that we can hope to know about reality.  This position stands in stark contrast to the representation of reality that emerges from the laws of physics that govern our everyday experience.  These laws are deterministic and allow a high degree of predictability.

The paper will explore a new approach to this problem that offers a possibility of eliminating the division.  The current solutions generally focus on the way that reality is.  The approach used in this paper is informed by the modern understanding that knowledge production is an active process and that the knower is an agent informs the approach taken by this paper.

The recognition of the agency of the knower is a relatively recent development.  In the past the agency of the knower was not in the frame of reference. Mental constructions were usually projected unconsciously and uncritically on reality that was regarded as given.[1] It was not before the beginning of the 20th century that physics, for example, recognized the agency of the knower. Einstein's theory of relativity and quantum mechanics are two good examples that illustrate this development. Theory of relativity proposed that there was no absolute space-time frame that all space-time frames were relative. It followed from this proposition that all frames depended on the choice of the knower and hence were equal. In the 1920s quantum mechanics boldly recognized that it was impossible to observe an object—say, a quantum system--without in any way disturbing it. Quantum mechanics stressed that all observations of quantum systems inevitably involve actions by the knower and the measuring devices used in experiments.

The recognition that the knower plays an active role in knowledge acquisition and that knowledge is a product of construction, not a mere reflection of reality, implies that what we know depends on the tools we use, including, most importantly, our mental tools. Consequently, what we know about how we know, how much we control our knowing has a decisive effect on our understanding of reality. It is from this perspective that the paper will approach the current controversy about the nature of reality. In contrast to the existing solutions that focus on the way that reality is, this paper will focus on issues related to the ways.

<u>Contemporary Science and Its Discontents</u>

Science has two faces. They represent two paths of our scientific endeavor. While one path seeks to understand reality, make it intelligible, the other pursues more operational, or instrumental goals and is interested more in results of experiments than in explaining the reality that lies behind these results. The former is commonly associated with what was known before the 19th century as "natural philosophy." The latter is mostly used in science today.

Although the two approaches are different, they are closely interrelated. Most scientists agree that rendering reality intelligible is not merely an intellectual exercise or a mental diversion, but rather helps to guide science in its more practical endeavors. As Peter Dear has concluded in his recent book:

> The intelligibility at the core of natural philosophy has never been inconsequential in the history of the sciences; instead, it has guided and shaped the very content of scientific knowledge, even while that knowledge has relied on appeal to instrumentality as an important complement to science's claim to provide true accounts of nature.[2]

Understanding reality is not a linear function of accumulating knowledge. There have been numerous instances in the course of human history when an expansion of our knowledge rendered reality less comprehensible, contradictory, and even paradoxical. Despite enormous progress in how much we know about reality that we have achieved, our current understanding of reality is very confusing.

---

[1] See, for example, a discussion of homocentrism and projection in Hooker, C. A. (1991). "Projection, Physical Intelligibility, Objectivity and Completeness: The Divergent Ideas of Bohr and Einstein." *Brit. J. Phil. Sci.*, 42, 491-511; Stapp, "Quantum Theory and the Role of Mind in Nature," p. 6; Folse, "Bohr's Conception of the Quantum Mechanical State of a System and Its Role in the Framework of Complementarity," pp. 4-6.

[2] Peter Dear, <u>The Intelligibility of Nature: How Science Makes Sense of the World</u> (Chicago: The University of Chicago Press, 2006), p. 174.

On one hand, there are many scientists who believe that reality is random and does not obey causal laws. For example, according to quantum mechanics, the processes that occur at the most fundamental level of reality—that is, at the level of elementary particles and atoms—are random and do not obey the laws of causality.[3] The universe described by quantum mechanics appears to make absolutely no sense when viewed outside its formalism. For example, how can one make sense of non-locality that involves speeds faster than the speed of light that is considered to be the absolute speed attainable in nature? Or, what should one make of superposition, according to which a quantum system can be in two different states at the same time. The contradictions with our familiar sense of how physical reality is are so great that even those who are intimately familiar with quantum theory find its puzzles hard to comprehend. Richard Feynman, who received a Nobel Prize for his achievements in quantum mechanics, cautioned:
> Do not keep saying to yourself, if you can possible avoid it, "But how can it be like that?" because you will get 'down the drain,' into a blind alley from which nobody has yet escaped. Nobody knows how it can be like that.[4]

Unsurprisingly, the view of quantum reality as random and uncertain sets limits to what we can know about it. In a characteristic remark Stephen Hawking, one of the most authoritative voices in modern physics, summarizes the view to which many contemporary physicists would subscribe:
> I do not demand that a theory correspond to reality because I don't know what it is. Reality is not a quality you can test with litmus paper. All I am concerned with is that the theory should predict the results of measurements. Quantum theory does this successfully.[5]

The view of reality as random is not limited the processes that occur at the level of elementary particles, or the micro level. Some physicists identify macro processes that display quantum phenomena. For example, a group of Russian physicists led by S. Korotaev has described phenomena of non-locality that occur in geomagnetic correlations.[6] Many biologists who subscribe to neo-Darwinism believe that the mechanism of the biological evolution involves random genetic mutations. The late Stephen J. Gould regarded contingency to be the basic creative force of life. In his view, contingency played a decisive role in the evolution: ". . . run the tape again, and the first step from prokaryotic to eukaryotic cell may take 12 billion years instead of two . . . ."[7]

---

[3] Here just some sources that advocate this view of reality: Geoffrey Hellman, "Einstein and Bell: Strengthening the Case for Microphysical Randomness," Synthese, vol. 53 (1982), pp. 445-60; Ulvi Yurtsever, "Quantum Mechanics and Algorithmic Randomness," arXiv: quant-ph/9806059 v 2, 13 December 2000 (accessed April 15, 2010)

[4] Online source at http://www.spaceandmotion.com/Physics-Richard-Feynman-QED.htm#Quotes.Richard.Feynman (accessed on October 20, 2008).

[5] Scientific American, July 1996, p. 65.

[6] S. M. Korotaev et al., "Experimental Study of Macroscopic Non-locality in Large-Scale Natural Dissipative Processes," NeuroQuantology, issue 4 (2005), pp. 275-94.

On the other hand, there are also many scientists who believe that reality is ordered and that a scientific description of this order is ultimately possible. Einstein's famous adage that "the Old One does not play dice" most succinctly summarizes this position. Although physicists apply deterministic perspectives mostly to the so-called macro domain, there are quite a few physicists and philosophers of physics who interpret quantum phenomena in terms of deterministic laws.[8] Determinists also challenge the contingency perspective on biological evolution. The famous biochemist Chrisitan de Duve advocates a deterministic interpretation of the origin of life on Earth,[9] as does Herbert Morowitz in his well-known book Beginnings of Cellular Life.[10]

These competing descriptions of reality are puzzling, and not just to laymen. Many physicists, for example, lament the lack of unity in contemporary physics that represents the physical universe as divided into two very different domains—the macro and the micro. The physicist Karl Svozil describes the situation in contemporary physics as nothing short of crisis.[11] Carlo Rovelli proclaims that the 20th century scientific revolution is "still wide open" since, in his view, ". . . our present understanding of the physical world at the fundamental level is in a state of great confusion." While recognizing great achievements of contemporary physics, Rovelli still thinks that both general relativity and quantum mechanics—the two most important theoretical perspectives in modern physics—"offer a schizophrenic and confused understanding of the physical world."[12]

---

[7] See S. J. Gould, Wonderful Life (London: Penguin Books, 1989), as quoted in Pier Luisi, "Contingency and Determinism" (Philosophical Transactions of the Royal Society of London A, vol. 361 [2003], p. 1142). Luisi echoes the same contingency view in his article: "At the present stage, one should accept the view that these few proteins of life are with us as the products of the bizarre laws of contingency, followed by chemical evolution processes" (ibid., p. 1144). See also Francois Monod, Chance and Necessity: Essay on the Natural Philosophy of Modern Biology (New York: Alfred A. Knopf, 1971) and Francois Jacob, The Possible and the Actual (Seattle: University of Washington Press, 1982).

[8] Hans Primas, "Hidden Determinism, Probability and Time's Arrow" in H. Atmanspacher and R. Bishop, eds., Between Chance and Choice: Inteerdisciplinary Perspective on Determinism (Thorverton: Imprint Academic, 2002), pp. 89-113 and Jean Bricmont, "Determinism, Chaos, and Quantum Mechanics," http://www.scribd.com/doc/11328575/Jean-Bricmont-Determinism-Chaos-and-Quantum-Mechanics (accessed September 12, 2009).

[9] C. de Duve, Blueprints for a Cell (Burlington, NC: Neil Patterson, 1991) and Life Evolving: Molecules, Mind, and Meaning (Oxford: Oxford University Press, 2002).

[10] H. J. Morowitz, Beginnings of Cellular Life: Metabolism Recapitulates Biogenesis (New Haven: Yale University Press, 1993).

[11] Karl Svozil's article "Science at the Crossroad Between Randomness and Determinism."

[12] Rovelli, "Unfinished Revolution," p. 1.

Although many physicists desire physics to provide a more coherent view of reality, the unification is not even in sight. Despite their aspirations for unity, most physicists, according to an article in the magazine <u>American Scientist</u>, continue to subscribe to the view that "the world is divided into two realms macro and micro, 'classical' and 'quantum,' logical and contradictory—or, as Bell put it in one of his essays, 'speakable' and 'unspeakable.'" What complicates the situation even more is that for many physicists "it is not clear where the border between the two realms should be . . .."[13]

The confusion as to the nature of reality raises unsettling questions: Is our world ultimately random or ordered? If it is random at the scale of elementary particles, how can be ordered and obey causal laws on the macro scales? And even more troubling questions lurk in the background: What ultimately is the use of our scientific enterprise if it cannot provide unambiguous answers about the nature of the world around us? What good is our knowledge about reality, if this reality is ultimately random and does not obey the laws of causality? How can we understand and control such reality? Is science the right direction to pursue in our understanding of reality or should we consider alternative ways of knowing?

This is not to overstate the case about the ultimate importance of these philosophical questions. Will the scientists stop doing science just because they have disagreements about the nature of reality? Few would believe that possible. However, making sense of reality has been and continues to be a major source of inspiration for our entire scientific enterprise. A belief that this goal is ultimately doomed to failure and that we may never know how reality actually is runs counter to the spirit that animates scientists and propels scientific progress.

It is not surprising that this climate of confusion has given rise to skepticism regarding our belief in the unlimited potential of science that has dominated our civilization at least since the beginning 19th century. The questioning of the scientific enterprise has already led to a certain "spiritualization" of science, "culture wars," "science wars," and an even greater confusion among both laymen and scientists. The awarding of the Templeton Prize in 2009 to the prominent French theoretical physicist and philosopher of science Bernard d'Espagnat is very symptomatic of the current climate in science. D'Espagnat was awarded the prize for "affirming life's spiritual dimension." The paradoxes of quantum theory have led d'Espagnat and, according to the magazine <u>Nature</u>, some other "serious scientists," to conclude that "reality, at its most basic, is perfectly compatible with what might be called a spiritual view of things."[14] In his book <u>On Physics and Philosophy</u>, as well as in his other writings, d'Espagnat argues that "reality is ultimately veiled from us," that science offers us "only a glimpse behind that veil," and that this reality is "in some sense divine."[15]  In his remarks to the Reuter's correspondent Tom Hanegan, d'Espagnat offered the following reflection:

> I believe we ultimately come from a superior entity to which awe and respect is due and which we shouldn't try to approach by trying to conceptualize too much.  It's more a question of feeling.[16]

---

[13] Nina Zanghi and Roderich Tumulka, "John Bell Across Space and Time," American Scientists (October 2003), viewed at http://www.americanscientist.org/bookshelf/pub/john-bell-across-space-and-time (accessed January 6, 2009).

[14] <u>Nature</u>, September 8, 2009.

D'Espagnat is not a lone voice in this "spiritualization."  His book <u>On Physics and Philosophy</u> in which he develops his views has been positively reviewed by some very prestigious publications.  The article in <u>Nature</u> cites several prominent scientists, including a neo-Platonist Roger Penrose, whose views resonate with those of d'Espagnat. [17]  To Antoine Suarez, the founding director of the Center for Quantum Philosophy in Zurich, contemporary theories related to quantum information suggest that "the physical reality is made of words [that] non-neuronal intellects speak to neuronal ones."  By contrast analogy with Laplace's and Maxwell's demons, he proposes to call these "non-neuronal intellects" quantum angels.[18]

The above is not to sound alarm about some tendencies in contemporary science.  After all, these "spiritual" tendencies are hardly prevalent among scientists and have limited effect on how they actually do science.  As the example of Newton and of other scientists shows, one can be a religious person and still do very good science.  Rather, this is to draw attention to the confusing descriptions that generate these tendencies.

However, this is also not to make light of the current confusion.  It is not entirely inconsequential either for the current cultural climate or for the progress of science.  Many scientists would like to see modern science produce a more coherent

---

[15] <u>Nature</u>, September 8, 2009.  The prestigious scientific magazine Science published an article about d'Espagnat under a disturbing title "Science Cannot Fully Describe Reality, Says Templeton Prize Winner" (<u>Science</u>, March 16 2009).

[16] Tom Hanegan, "French Physicist d'Espagnat wins Prestigious Templeton Prize, Reuters, September 16, 2009 at http://www.reuters.com/article/idUSTRE52F2GC20090316 (accessed July 17, 2010).  D'Espagnat's comments for <u>The Times</u> (June 16, 2009) were very much in the same vein. "Mystery," he said, "is not something negative that has to be eliminated.  On the contrary, it is one of the constitutive elements of being."

[17] <u>Nature</u>, September 8, 2009.

[18] Antoine Suarez, "Classical Demons and Quantum Angels.  On 't Hooft's Deterministic Mechanics" (arXiv:  0705.3974 v.1 [quant-ph] 27 May 2007, accessed April 29, 2009), pp. 6 and 13.   Elsewhere in the same article, Suarez further elaborates:  "In conclusion, the experiments testing quantum entanglement rule out the belief that physical causality necessarily relies of observable signals and that an observable event (the effect) always originates from another observable event (the cause) occurring before in time.  This means that quantum correlations have roots outside space-time and, in this sense, originate from a free and intelligent agent.  One is led to accept 'the two freedoms':  the freedom of the experimenter and the freedom of Nature, and to see quantum randomness as a particular expression of free will" (Suarez, Classical Demons, p. 6)

understanding of reality.  They believe that a more unified vision will offer better prospects for the future development of science.  John Baez reflects this attitude in the following passage:
> General relativity and quantum field theory are based on some profound insights about the nature of reality.  These insights are crystallized in the form of mathematics, but there is a limit to how much progress we can make by just playing around with this mathematics.  We need to go back to the insights behind general relativity and quantum field theory, learn to hold them together in our minds, and dare to imagine a world more strange, more beautiful, but ultimately more reasonable than our current theories of it.[19]

Expressing a wide-spread hope in the scientific community, Adrian Kent writes:  ". . . almost everyone suspects that a grander and more elegant unified theory . . . await us."[20]  However, despite these passionate appeals and hopes, the goal of having a unified picture of reality continues to be elusive and the question of whether reality is random or deterministic remains unresolved.

Defining Randomness and Determinism

Despite numerous attempts to define randomness and its antipode, determinism, many who made such "heroic efforts" recognize that these concepts are not easy to define.[21]  Commonsense definitions are deceptively simple.  Definitions of randomness, for example, usually focus on such characteristics as the absence of any noticeable pattern, regularity, or constraint.  Oxford English Dictionary defines randomness as "having no definite aim or purpose; not sent or guided in a particular direction; made, done, occurring, etc. without method or conscious choice; haphazard."  Wikipedia provides a similar description of randomness as "a concept of non-order or non-coherence in a sequence of symbols or steps, such that there is no intelligible pattern or combination."

However, on close analysis one finds both these definitions problematic.  For example, how can we prove that there is no pattern to a given phenomenon?  Is the sample large enough?  Are there constraints of which we are not aware?  These are just some of the questions that point to the difficulties of defining randomness.

Despite definitional difficulties, randomness and determinism are so important for contemporary science that scientists still have to come up with some ways of talking about them.  Thermodynamics, for example, associates randomness with its central concept of entropy.  As used in thermodynamics, randomness is the amount of energy unavailable for doing work.  In another current definition, that is used in several scientific disciplines, including statistics, randomness is the state of a dynamic system in which all possible outcomes are equally probable; in other words, randomness is defined as equal probability.  Finally, there is a definition that is broadly used in mathematics and computer science and is associated with the Kolmogorov-Chaitin complexity. In this definition, a random string of numbers is a string that has no shorter description than that string itself.[22] Computer scientists also use the terms "incompressibility" or "irreducibility" in association with this understanding of randomness.[23]

---

[19] As quoted in John Small, "Why do Quantum Systems Implement Self-Referential Logic?  A Simple Question with a Catastrophic Answer," in D. M. Dubois, ed., Computing Anticipatory Systems:  CASYS'05:  Seventh International Conference (American Institute of Physics, 2006), p. 167.
[20] Kent, Night thoughts, p. 77.
[21] See Gina Kolata, "What Does it Mean to Be Random?"  Science, Vol. 231, No. 4742 (March 7, 1986), p. 1068.

Randomness and determinism are also frequently associated with two other very common concepts in science—equilibrium and disequilibrium.[24] The abovementioned definitions of randomness as equal probability or as incompressibility of a string of numbers reflect this common association. An equal probability of different states in a system indicates that the system is in equilibrium. The definition of randomness as incompressibility and irreducibility, widely used in mathematics and computer science, emphasizes uniqueness of a random string or a state; that is, it indicates that a string or a state cannot be traced to any other known string or state. In other words, such string or state has no cause. The absence of a cause, or acausality, signals the presence of equilibrium since fully equilibrated states have no time arrow, no before or after, and hence no causality.

It is very common, on the other hand, to connect determinism, with entropy production and the presence of the time arrow pointing toward the future.[25] Both production of entropy and the time arrow are only possible if the initial level of entropy is low, that is, if the system is in disequilibrium. The best-known example of the association of determinism with disequilibrium is the currently dominant cosmological theory on the origin and evolution of our universe commonly known as Big Bang. According to this theory, our universe began with the initial expansion when entropy was low and the universe was in disequilibrium. After the initial expansion the universe has been continuing to evolve in accordance with the Second Law of Thermodynamics towards the final equilibrium, or the so-called "thermal death." Thus our universe inevitably, one could say deterministically, evolves towards this grim future.

---

[22] See, for example, G. J. Chaitin, "Randomness and Mathematical Proof," Scientific American, No. 232, pp. 47-52.

[23] See, for example, Stephen Wolfram, A New Kind of Science, available online at http://www.wolframscience.com/nksonline/toc.html (accessed on May 3, 2009), pp. 554-55 and Chris Thorn, "Randomness and Entropy," (SANS Institute, 2003), p. v.

[24] See, for example, M. Grender, Jr. and M. Grender, "Randomness and Equilibrium. Potential and Probability Density," in L. R. Fry, ed., Bayesian Inference and Maximum Entropy Methods in Science and Engineering: 21st International Workshop (Melville, New York: American Institute of Physics, 2002), vol. CP617, pp. 405-10; also A. Sengupta, "Is nature quantum non-local, complex, holistic, or what?" I: Theory and Analysis, Non-linear Analysis: Real World Applications (2008).

[25] See, for example, Hans Reichenbach, The Direction of Time (Berkeley: California University Press, 1956) who associates the time arrow with causality and entropy production.

One can also encounter in scholarly literature differentiations of randomness and determinism as ontic or epistemic. Another differentiation along a somewhat similar line is between pseudo and real, or sometimes true. There are also other differentiations. Subnik Chakraborty, for example, identifies as many as four kinds of randomness: ontic, epistemic, pseudo and telescopic.[26] An assessment of the validity of these differentiations is beyond the scope of this paper. However, at the risk of oversimplification I will use the basic differentiation between ontic and epistemic: the former having to do with the way things are and the latter with the ways we know.[27]

This differentiation reflects an issue that is critical to this paper: Is randomness or determinism intrinsic to nature or are they merely due to the imperfection of our knowledge? It is not difficult to see that this question goes right to the heart of the main problem that this essay addresses: Is reality random or deterministic?

Epistemological Roots of the Problem of Randomness vs. Determinism

As has been indicated earlier, there are two principal positions on the nature of reality among scientists. According to one position, reality as ultimately random and unpredictable; the other position views reality essentially in deterministic terms. The first position largely owes its inspiration to quantum mechanics. Francis Bailly and Giuseppe Longo, for example, relate randomness to quantum non-separability and non-locality, and regard it as intrinsic to the processes that occur on the level of elementary particles—the level that they, among many others, consider the most fundamental to nature.[28] Geoffrey Hellman in his piece "Einstein and Bell: Strengthening the Case for Metaphysical Randomness" makes a similar argument in support of the ultimately random behavior of quantum mechanical systems.[29] Others, like Jean Bricmont and Hans Primas, see ontic determinism lurking behind the appearance of quantum randomness.[30]

---

[26] Chakraborty, "On Why and What."
[27] One could note that this differentiation reflects the influence of traditional dualism in which mind and matter are diametrically opposed to each other. As I have argued elsewhere, dualism is not a product of empirical observation; rather, it is a necessary outcome of the epistemological perspective that does not recognize a vital and constructive link between the subject and the object. Obviously, if the process that constructs both the subject and the object is not in the knower's field of vision, the two appear as ontologically separate and opposed to each other (see Gennady Shkliarevsky, "Deconstructing the Quantum Debate: Toward a Non-Classical Epistemology," arXiv:0809.1788v1 [physics.hist-ph], p. 8).
[28] Francis Bailly and Giuseppe Longo, "Randomness and Determination in the Interplay Between the Continuum and the Discrete" in Mathematical Structures in Computer Science, Vol. 17, Issue 2 (April 2007), pp. 289-305.
[29] Geoffrey Hellman, "Einstein and Bell: Strengthening the Case of Microphysical Randomness," Synthese, Vol. 53 (1982), pp. 445-60.
[30] Jean Bricmont, "Determinism, Chaos, and Quantum Mechanics," http://www.scribd.com/doc/11328575/Jean-Bricmont-Determinism-Chaos-and-Quantum-Mechanics (accessed June 22, 2010); Hans Primas, "Hidden Determinism,

Despite the fact that the two positions are diametrically opposed to each other, they do share some unsettling questions: If you have some random or deterministic phenomena, how do you know that they are truly random or truly deterministic? Can one demonstrate that the randomness or determinism of these phenomena is truly ontic?

In his article Ulvi Yurtsever makes a strong argument that quantum mechanical probabilities are truly genuine, that is, that they are algorithmically random, or incompressible. However, he also emphasizes that "no algorithmically incompressible binary string can ever be <u>constructed</u> via a finitely-prescribed procedure (since, otherwise, such a procedure would present an obvious algorithm to compress the string thus obtained)."[31] This observation recognizes that although truly algorithmically random strings may indeed exist, their existence cannot be demonstrated.

In the opposite camp, Jean Bricmont's analysis yields a result that simply dismisses the entire issue of the intrinsic nature of determinism as ultimately irrelevant. Bricmont examines two current definitions of determinism. He finds that one definition in which determinism is conflated with predictability renders determinism trivially false. As to the other definition that avoids conflation, Bricmont raises a question whether there is a function--in a Platonic sense (that is, independent of our ignorance)--that determines a finite sequence of sets of numbers that never repeats itself in a unique way. His answer is that the existence of such function is simply impossible to disprove because one can always find a function or even many functions that map "each set into the next one."[32] Bricmont's conclusion dismisses the whole issue of determinism as utterly irrelevant to science. In his view, "there is no notion of determinism that would make the question [of determinism] scientifically relevant . . . ontically it [determinism] is true but uninteresting [that is, impossible to disprove]."[33] "I don't know," he adds, "how to formulate the issue of determinism so that the question becomes interesting."[34]

For Hans Primas, determinism refers strictly to ontic descriptions. Like Bricmont, he makes a very convincing argument against conflating, as is often done, determinism with predictability. Even quantum interactions, he stresses, which are notoriously unpredictable, are "governed by <u>strict</u> statistical laws."[35] Primas follows the principle of scientific determinism as formulated by the French mathematician Jacque Hadamard. According to this principle ". . . in a well posed forward-

---

Probability, and Time's Arrow," in H. Atmanspacher andR. Bishop, eds., <u>Between Chance and Choice. Interdisciplinary Perspectives on Determinism</u> (Thorverton: Imprint Academic, 2002), pp. 89-113 (accessed through online version at http://philsci-archive.pitt.edu/archive/00000948/ on August 12, 2010).

[31] Ulvi Yurtsever, "Quantum mechanics and Algorithmic Randomness," arXiv:quant-ph/9806059v2 13 Dec 2000 (accessed May 14, 2008), p. 1.

[32] Bricmont, p. 4.

[33] Bricmont, p. 4

[34] Bricmont, p. 1.

[35] Primas, <u>Hidden Determinism</u>, p. 1 (emphasis in the original).

deterministic dynamical system every initial state determines all future states uniquely."[36] However, in contrast to others that subscribe to similar definitions of determinism (for example, Laplace), Primas follows Hadamard in regarding the principle of determination as regulative, and not in some absolute sense; in other words, if in some cases this principle is not satisfied, "it can be enforced by choosing a larger state space."[37] According to Primas, such enforcement is perfectly compatible with mathematical probability theory because:

> Every mathematically formulated dynamics of statistically reproducible events can be extended to a description in terms of a one-parameter group of automorphisms on an enlarged mathematical structure which describes a <u>fictitious hidden determinism</u>. Consequently, randomness in the sense of mathematical probability theory is only a weak generalization of determinism.[38]

It is not difficult to see similarities in the way that Bricmont and Primas resolve the problem of determinism. Both see that by enlarging the state space one can always find a deterministic function for a sample or a set. This solution resonates with the famous proof of consistency and completeness by the Austrian logician and mathematician Kurt Godel. As Godel has shown, any deductive system can have true sentences whose truth is indemonstrable. In order to demonstrate their truth, one should resort to meta-mathematical procedures and construct a new and broader axiomatic structure that would be powerful enough to make such proof possible. However, according to Godel's proof, even the new and enlarged structure will not be able to escape the same paradox as it will also allow other true but improvable sentences.[39]

As one can see from the above, the three authors have essentially reformulated the whole problem of randomness vs. determinism. In the new formulation, the problem is no longer whether randomness or determinism objectively exist, but rather whether one can offer a proof of this existence. Thus they transform the problem from ontological into epistemological, or from how reality is to how we know. The connection, whether explicit (Bricmont) or implicit (Yurtsever and Primas), with Godel is also very indicative and significant insofar as Godel's proof deals with how we know. If the solution of the problem of randomness vs. determinism lies through epistemology, as the above interpretations suggest, it is logical to propose that its origin may also lie in how we know rather than in what is out there.

One can also glean the connection of this problem to epistemology from another angle. There is a great deal of empirical evidence suggesting that nature does not give preference to either randomness or determinism. In fact, many natural phenomena point to a close relationship and complex interaction between random and deterministic processes.  Many processes in nature can be often classified as random and deterministic at the same time.[40]  The Nobel laureate Ilya Prigogine noted a close relationship between random and deterministic processes in his book with a characteristic title Order out of Chaos.[41] In his best selling book A New Kind of Science Steven Wolfram also shows that randomness can evolve into order and vice versa.[42] Adducing to the fractal geometrical patterns in nature, Paul Carr observes that many natural phenomena reveal "the complex interplay between randomness (symbolized by dice) and global determinism (which loads the dice).  The Neo-Darwinist approach to evolution as Carr points out, also emphasizes interplay between random genetic mutations and the globally deterministic natural selection.[43]  Summarizing the evidence related to such diverse phenomena as turbulent flows and neurons, Tamas Viscek in his article that appeared in Nature stresses that:

> . . . in both these systems [turbulent flows and neurons] (and in many others), randomness and determinism are both relevant to the system's overall behavior.  Such systems exist on the edge of chaos, they may exhibit almost regular behavior, but also can change dramatically and stochastically in time and/or space as a result of small changes in conditions.[44]

---

[36] Primas, <u>Hidden Determinism</u>, p. 10.
[37] Primas, <u>Hidden Determinism</u>, p. 10.
[38] Primas, Hidden Determinism, p. 1 (emphasis in the original).
[39] See Ernest Nagel and James R. Newman, <u>Godel's Proof</u> (New York: University Press, 1953).

In another piece, also published in Nature, Kees Wapenaar and Roel Snieder make a  similar point, drawing on evidence from physics:

> Our view of the universe may have shifted from the deterministic to the random, but since the turn of the last century physics itself has provided a less simplistic view.  Fields generated by random sources can be used for imaging and for monitoring of systems such as Earth's subsurface, or mechanical structures such as bridges.  Randomness is no longer at odds with determinism, it has instead become a new window on the deterministic response of the physical world.[45]

As the physicist Joseph Ford succinctly put it,  "God plays dice with the universe.  But they are loaded dice."[46]

There have also been challenges to the exclusive emphasis on randomness central to standard quantum mechanics.  In the most recent one, the physicists Sheldon Goldstein, Detlef Dürr, and Nino Zhangi offer an interpretation of quantum mechanics that is, in Goldstein's words, "precise,

---

[40] Berkowitz, et al., "Ergodic Hierarchy," p. 661.
[41] I. Prigogine and  I. Stengers, Order out of Chaos (New York:  Bantam Books, 1984), particularly pp. 292-95.
[42] Steven Wolfram, A New Kind of Science (2002)
[43] Paul H. Carr, "Does God Play Dice?  Insights from the Fractal Geometry of Nature," Zygon, vol. 39, no. 4 (December 2003), p. 934.
[44] Tamas Vicsek, "The Bigger Picture," Nature, Vol. 418 (11 July 2002), p. 131.
[45] Kees Wapenaar and Roel Snieder, "Determinism:  ," p. 643.
[46] James Glieck, Chaos:  Making a New Science (New York:  Penguin, 1987), p. 314

objective—and deterministic."[47] In their view, the observed randomness is merely apparent. In another challenge, the data obtained in the study of neutron resonances have led a group of physicists at Oak Ridge Electron Linear Accelerator, headed by Dr. Paul Koehler, to question the applicability of random matrix theory to movements of neutrons and protons in the nucleus. The data indicate that the particles in the nucleus are moving in a coordinated fashion, rather than randomly as suggested by random matrix theory.[48] At the same time other physicists report observing quantum phenomena in macro events. A group of Russian physicists, led by S. M. Korotaev, has observed the phenomenon of non-locality, usually associated with the quantum domain, in dissipative geomagnetic macro processes.[49]

Empirical evidence also shows that nature does not favor either equilibrium (associated with randomness) or disequilibrium (associated with determinism). For example, in his interpretation of the current state of the universe, the astrophysicist Manasse Mbonye conjectures that "the universe is always in search of a dynamical equilibrium," which suggests an interplay between the states of equilibrium and disequilibrium.[50] Although the currently dominant cosmological theory asserts that our universe originated in the state of original disequilibrium, or Big Bang, numerous critics of this theory point to its speculative nature and argue that since it is an extrapolation from the current conditions into the past, this theory is not justified and still lacks unambiguous empirical support.[51]

Why, then, in view of this substantial evidence to the effect that reality shows no preference for either randomness or determinism, the current solutions of the problem are one-sided? Or, rather, what does the fact that these solutions are one-sided tells us?

It is obvious that the selection and interpretation of facts in the current solutions favors a one-sided interpretation. Since the bias toward one-sidedness does not occur in one isolated case, one cannot invoke ignorance as a possible explanation. Rather, one can suggest there are powerful factors at play that determine these choices. And these factors must be subjective in nature, that is, they are not due to the way reality is, but rather to the way of knowing.

---

[47] Buchanan, "Quantum Randomness,"
[48] See "Nuclear Theory Nudged," Nature, Vol. 466, No. 7310 (August 26, 2010), p. 1034.
[49] S. M. Korotaev, et al., "Experimental Study of Macroscopic Non-locality of Large-Scale Natural Dissipative Processes," NeuroQuantology, Issue 4 (2005), pp. 275-94.
[50] Manasse Mbonye, "Constraints on Cosmic Dynamics," arXiv:gr-qe/0309135v1 30 Sep 2003, pp. 1-2 (accessed November 21, 2008).
[51] Sean Carroll, for example, observes that ". . . scenarios of this type are extremely speculative and may very well be wrong" (Sean Carroll, "Is Our Universe Natural?" arXiv:hep-th0512148v1 13 Dec 2005, p. 5 (accessed February 21, 2010). Paul Steinhard and Neil Turok—two prominent critics of Big Bang—also point to the speculative nature of this theory and counter it with their own cyclical theory of the universe (Paul J. Steinhard and Neil Turok, "A Cyclic Model of the Universe," Science, Vol. 296, Issue 5572 (May 24, 2002), pp. 1436-40.

Randomness, Determinism, and the Nature of the Real

Despite the radical differences between the two principal positions on the problem of randomness vs. determinism, there is one fundamental aspect that they share.  They both cannot bring randomness and determinism together.  In both cases, the epistemological perspective does not permit such integration.  As a result, each position has to make a choice of either determinism or randomness, but not both.

The need to make a choice and the failure to provide an interpretation that would be capable of integrating the two opposites reveal an enduring influence of the traditional conception of knowledge.  This conception views knowledge as a product of more or less passive reflection and does not recognize in any significant way the process by which knowledge is produced.  As has been indicated earlier, the failure to recognize the process of construction of knowledge results in dualism, or viewing reality in terms of irreconcilable binaries.  As I have argued elsewhere, [52] the most enduring binary opposition—that of the subject and the object—did not result from empirical observations.  Rather, it was a necessary outcome of the epistemological perspective that simply did not see the vital link—the process of construction—that connected the subject and the object.   As a result, the two appear as disconnected and opposed to each other.  Many other constructed binaries—such as mind-matter or randomness-determinism—can be traced to this fundamental failure to recognize the process of the construction of knowledge and to understand its role.[53]  Incidentally, the current philosophical differentiation between instrumental science and natural philosophy[54] seems to be equally unjustified as it implies a distinction between thinking and doing, as if thinking is not doing.

Classical epistemologies did not recognize the agency of the knower and did not see this link.  In their conception, knowledge appeared as a mere reflection of reality and the knower as a more or less passive observer who was deemed to be capable of observing reality without in any way disturbing it.  Simply disregarding the agency of the knower certainly did not eliminate its vital role.  It merely allowed to project mental constructions unconsciously and uncritically on reality, and to substitute these projections for reality.[55]

---

[52] See, for example, Peter Dear, <u>The Intelligibility of Nature:  How Science Makes Sense of the World</u> (Chicago:  The University of Chicago Press, 2006).

[53] See Gennady Shkliarevsky, "Deconstructing the Quantum Debate:  Toward a Non-Classical Epistemology, arXiv:0809.1788v1 [physics.hist-ph], p. 8, and "Of Cats and Quanta:  Paradoxes of Knowing and Knowability of Reality" (unpublished manuscript under review), pp. 9-10.

[54] See, for example, Peter Dear, <u>The Intelligibility of Nature</u>.

[55] See, for example, a discussion of homocentrism and projection in Hooker, C. A. (1991). "Projection, Physical Intelligibility, Objectivity and Completeness:  The Divergent Ideas

It has been well over a century since a radical departure from the classical view of the role of the knower.  The pivotal point in this departure was the recognition that the knower plays a vital role in constructing knowledge.  Two very important examples of this innovation in science were Einstein's theory of relativity and quantum mechanics.  The introduction of the point-of-view invariance for the frame of reference was seminal for the theory of relativity.  In Einstein's view, space should look invariant regardless of the frame chosen by the knower. Einstein's dictum was that no frame should be given preference.  This central tenet contained a powerful recognition that all frames are constructed and therefore all are equal.  The only non-relativistic component in Einstein's picture of the universe was light.  The speed of light had to be the same for all frames, and therefore constant.  If it were not, then some frames had to be different from others, which would contradict Einstein's principal tenet.

Quantum theory was even more radical in its recognition of the agency of the knower.  It no longer viewed the knower as a passive observer but rather as an active agent whose interaction with a quantum system could change it.  According to quantum theory, the knower's choices, most importantly what and how to measure, radically affected, one could even say produced, the outcome of experiments (for example, measurements performed on a particle).  The legendary physicist John Wheeler probably best exemplified this radicalism in his comment that the cosmos "has not really happened, it is not a phenomenon until it has been observed to happen."

Once the agency of the knower was recognized, the acceptance of the notion of the constructed nature of knowledge was soon to follow, and with it, the interest in the process of construction.  The early pioneering studies of the construction of knowledge, such as the works of Jean Piaget, paved the way for a growing number of contemporary interdisciplinary theoretical approaches that focus on the process of construction, such as systems theory, communication theory, theory of emergence, constructionism, chaos theory, complexity theory, theory of autopoiesis, as well as more disciplinary fields such as evolutionary and developmental psychology, evolutionary and developmental biology, economics, management science, and others.

The focus on the process of construction has brought forth a new epistemological perspective that is decidedly non-dualistic.  Piaget, for example, has shown that any advance in understanding reality by the child necessarily involves changes in the child's mind and vice versa.  In this perspective, the subject and the object no longer stand opposed to each other but are engaged in a productive and mutually enriching relationship.  Only when we disregard the process of construction, the two appear as diametrically opposed to each other.

Piaget's studies have also shown that the process of the construction of knowledge is characterized by dynamic equilibrium.  Dynamic equilibrium

---

of Bohr and Einstein," <u>Brit. J. Phil. Sci.</u>, 42, 491-511; Stapp, "Quantum Theory and the Role of Mind in Nature," p. 6; Folse, "Bohr's Conception of the Quantum Mechanical State of a System and Its Role in the Framework of Complementarity," pp. 4-6.

involves a balance between equilibration and disequilibrium. Any increase in equilibrium necessarily involves at the same time an equivalent increase in disequilibrium and vice versa.[56] In his study of the origin of conscious intellect in children,[57] Piaget demonstrates how the equilibration of reflex functions—such as seeing or hearing—generates operations that are more powerful than reflex functions (for example, operations that are capable of constructing permanent mental images). This power differential creates an imbalance commensurate with the increase in equilibrium among reflex functions.

One can see in this example that equilibrium and disequilibrium no longer appear as independent states diametrically opposed to each other and mutually exclusive. Rather, they emerge as intimately related aspects of the same process. Equilibration leads to the growth of equilibrium and disequilibrium at the same time at the two adjacent but very different levels of organization. Equilibration gives rise to common operations that regulate entities involved in the process of equilibration. These operations are more powerful than each individual entity or their sum total. They are more powerful because their combinatorial potential is much higher than that of the individual entities they regulate. Using the example of Piaget's study mentioned above, the equilibration of functional operations, such as hearing and seeing, offers the combinatorial possibility of seeing when hearing and hearing when seeing. The enhanced combinatorial power does not stop there. The combination of hearing and seeing leads to the emergence of permanent mental images that are present even when actual objects are not. Such mental images open the path to symbolic operations that are practically unlimited in their combinatorial capacity.

Incidentally, the emergence of new properties associated with greater combinatorial power requires rethinking such fundamental concept as causality. The current understanding of causality, shaped by traditional epistemology with its dualistic approach, defines relations as causal when one can reduce—in other words, explain—a state of a system to either the interaction of its components (spatial reductionism) or to another state that precedes it in time (temporal reductionism). One can easily see the inadequacy of this conception of causality when applied to dynamic systems and processes. We know that a system originates from local interaction of components that eventually become its subsystems. However, can we reduce the former to the latter? We certainly cannot. The combinatorial capacity of a system is far more powerful and extensive than those of its subsystems. It is certainly not possible to reduce something that is more powerful to something that is less. Also, the interactions of components that generate systemic constraints certainly precede the emergence of a system in time

---

[56] See, for example, Jean Piaget, <u>The Equilibration of Cognitive Structures: The Central Problem of Intellectual Development</u> (Chicago: The University of Chicago Press, 1985), particularly pp. 10-15.
[57] Piaget, The Origin of Intelligence in Children (etc. ).

but, again, for the same reasons as stated above the latter cannot be reduced to the former.

I have already mentioned earlier the connection that is often made between equilibrium and disequilibrium, on one hand, and randomness and determinism, on the other. Since randomness implies equal probability of all possible interactions without any distinct path and with no time arrow, it is certainly an intrinsic property of equilibrium. A characteristic feature of disequilibrium is a distinct path of interactions that necessarily, one could also say deterministically, leads from unequal probability to equal probability. This unique path allows a clear differentiation between before and after and, consequently, has the time arrow that points toward the future. There is a clear distinction in this unique path between before and after and consequently the arrow of time pointing toward the future. The presence of the arrow of time is a necessary condition for causality and determinism.

The fact that equilibrium and disequilibrium, on one hand, and randomness and determinism, on the other, are closely related suggests that, just like equilibrium and disequilibrium, randomness and determinism also do not exist independently of each other and neither is dominant over the other. The two are always in balance. Only when we view reality from a perspective that does not take into account the process of construction, we see equilibrium and disequilibrium as two separate and diametrically opposed states. By the same logic, the properties that characterize these states—randomness and determinism—only appear to us as separate and diametrically opposed when we view reality from a limited perspective of either equilibrium or disequilibrium but not both. And just as in the case of equilibrium and disequilibrium, the reason why randomness and determinism can coexist with each other without creating a paradox is the fact that they are both part of the same process that functions simultaneously at two different, albeit adjacent, levels of organization. What may appear as random when viewed from the one level of organization will appear as perfectly ordered when viewed from another level of organization. For example, interactions of the cells in an organism may appear chaotic and unpredictable if more powerful global constraints of the organism that regulate the behavior of the cells are not taken into consideration.

So far the focus of this discussion has been mostly on the construction of knowledge. Although the construction of knowledge is an important part of what is going on in the universe, it certainly cannot stand for reality as a whole. However, if we are to take the notion of evolution seriously, we must conclude that construction of knowledge cannot be an isolated process that has nothing to do with other processes that are taking place in the universe. Our ability to construct knowledge could not have appeared out of nowhere; it could have only emerged from the processes that preceded it in the course of the evolution. As has been pointed out earlier, Piaget has shown the process of the emergence of mental operations from the equilibration of physiological functions of the organism, thus pointing to a link between psychological functions and biological and chemical processes.[58]

As this paper has already pointed out, the representation of mind and matter as diametrically opposed has been largely due to the disregard of the process of construction that links the subject, or how we think, and the object, or the way things are. The process of construction is not unique to the human race. Humans have inherited it in the course of the evolution and transformed into a powerful tool for their advancement. As a product of the evolution, the construction of knowledge is but a particular case of the more general process of organization and creation of new forms that we observe at all levels of reality. Therefore, the two must share dynamic features that make them possible, such as dynamic equilibrium, or the balance between the processes of equilibration and disequilibrium.

And there is evidence that they actually do. Just like we find dynamic equilibrium in the construction of knowledge, one can observe the interplay between equilibrium and disequilibrium, or randomness and determinism, in the processes that take place at many other levels of organization of reality: from the sub-atomic level all the way to the cosmic scale. The astrophysicist Manasse Mbonye, for example, sees the interplay of equilibrium and disequilibrium in the processes of space expansion and the creation of matter in our universe. In his view, "the universe is always in search of a dynamical equilibrium."[59] The physicist Paul Carr, echoing the ideas of Stuart Kaufman, also sees interplay between randomness and determinism as "the basis of the inherent creativity of the natural order and its ability to generate new forms of matter and life."[60] Kees Wapenaar and Roel Snieder offer the following generalization in their article that appeared in the magazine <u>Nature</u>:

> Our view of the universe may have shifted from the deterministic to the random, but since the turn of the last century physics itself has provided a less simplistic view. Fields generated by random sources can be used for imaging and for monitoring of systems such as Earth's subsurface, or mechanical structures such as bridges. Randomness is no longer at odds with determinism; it has instead become a new window on the deterministic response of the physical world.[61]

The ubiquity of dynamic equilibrium in nature suggests that randomness and determinism, just like equilibrium and disequilibrium, do not exist on their own. They are closely interrelated aspects of the general process of organization of reality. They only appear as separate and diametrically opposed when we abstract them from this process. Interactions among subsystems in a system may appear random if we do not take into account global regulations; when these global regulations enter into the field of vision, they will appear perfectly ordered. "Does

---

[58] Piaget,

[59] Manasse Mbonye, "Constraints on Cosmic Dynamics," arXiv:gr-qe/0309135v1 30 Sep 2003, pp. 1-2 (accessed November 21, 2008).

[60] Carr, p. 934.

[61] Wapenaar, "Determinism," p. 643.

God play dice?" Paul Carr asks, "Yes and no. Yes, if one considers the random nature of evolution and fractal statistics. No, if one considers their globally deterministic laws and rules."[62]

To recapitulate, randomness and determinism are not separate states in some absolute ontological sense. They only appear so as abstract idealizations of real conditions encountered in nature. They are the ways that these conditions appear to us when we view them from a limited perspective that does not take into account the process of construction. For example, when viewed from the perspective of interacting subsystems--that is, taking no account of the regulatory system—reality appears to be random and chaotic. When, however, viewed from the perspective of the system that regulates the interaction of subsystems, reality appears to be deterministic.

These two limited perspectives correspond to the two approaches in studying reality that are currently in use. The first perspective that emphasizes the interaction of subsystems constitutes the core of the atomistic reductionist approach. The second perspective that focuses on regulatory systems makes up the basis for the holistic approach. The atomistic reductionist approach is by far the more popular of the two. It is important to stress, however, that both approaches are limited and inadequate for providing a comprehensive description of reality. The atomistic approach fails because it tries to reduce the system to its subsystems and their interactions. Such reductionist explanation should fail by definition because it is impossible to reduce (explain) a system to the subsystems that it regulates; the latter are simply not powerful enough. The holistic approach also fails to provide a comprehensive description. It does not and cannot show how a system originates. The system either simply exists or appears as if by a miracle from an unknown intelligent source.

The inadequacy of these two approaches shows that an understanding of the dynamic evolution of reality from one level of organization to another is essential for providing a comprehensive description. There is a growing realization of the need to abandon the old atomistic, reductionist perspective. It has certainly been successful but the limitations of its one-sidedness are becoming increasingly evident. However, replacing it with its antipode—a holistic approach—as some suggest,[63] will certainly not do. A holistic approach will be just as blind to the process of construction as the traditional reductionist atomism has been. The only way to advance toward a fuller understanding of reality that would avoid the pitfalls of dualism is to focus our attention on the process of construction. As decidedly non-dualistic, this approach brings randomness and determinism together and explains their close interrelation that one can widely observe in nature.

There is still much that we should know about this dynamic evolution; many details and lacunas still need to be filled. But some general contours are already emerging. The perspectives that focus on the ways that reality organizes itself and creates its new forms of show how local interactions among subsystems give rise to global systemic constraints, or, in other words, how a system emerges from the interaction of the components that become its subsystems, and how the newly emerged system regulates the interaction of subsystems.

---

[62] Carr, Does God Play Dice?, p. 937.
[63] See Mendel Sachs

This paper has argued that reality is neither orderly nor chaotic. What we see as two separate states are merely aspects of the process of constant organization. Reality is always moving, always in the process of transition from one level of organization to another. For example, what we posit as the initial state of order—Big Bang—in which our universe supposedly originated has more to do with the unconscious epistemological preferences than we actually are willing to admit. We may very well be asking in vain questions about the source of this initial order. The view of reality as constantly moving from one level of organization to another suggests that there is no such initial state but merely an infinite cycle of states following one another. One can pose a legitimate question in relation to the theory of Big Bang that posits such initial ordered state: Can we really reduce the current state of the universe to those that preceded it in time, particularly by 14 billion years? By the same token, is it possible to extrapolate the future state of the universe from the current one? Can these subsequent states that have much greater combinatorial power be reduced to the preceding less powerful ones?

The view of the universe that eternally evolves from one level of organization to another, rather than from a highly ordered state to the final thermal death, the view where order and disorder, randomness and determinism do not stand opposed to each other but are merely aspects of the process of this evolution does not reject the Second Law of Thermodynamics that serves as the foundation of the current cosmological theories associated with Big Bang; it merely requires rethinking of how this law is interpreted and applied. The Second Law says that in any closed system entropy production will either increase or will be 0. In accordance with the new perspective, equilibration, or the growth of entropy, at one level of organization of reality will always create disequilibrium, organization, and consequently a decline in entropy at another level. If the two are balanced, the total entropy of the universe will always be 0.

As this paper has shown, the problem of randomness vs. determinism is not intrinsic to reality. Rather, it is created by one-sided approaches, either atomistic reductionist or holistic, that takes no cognizance of the fundamental dynamic evolution that pervades reality and constitutes its most defining feature. This paper has shown the ultimate futility of trying to define reality in a one-sided manner as either random or deterministic. It is neither. Randomness and determinism, and their concomitant states or equilibrium and disequilibrium, are merely abstractions that make their appearance when we approach reality in a one-sided manner. Reality is eternally balanced between equilibrium and disequilibrium. It is this balance that makes its dynamic evolution possible. Reality never stands still but constantly evolves from one level of organization to another. It is neither in equilibrium nor in disequilibrium, it is neither random nor chaotic, but rather it always exists, as Stuart Kauffman aptly put it, "at the edge of chaos."